# Chip-integrated metasurface-enabled single-photon skyrmion sources

Xujing Liu[1], Yinhui Kan[1,2]*, Shailesh Kumar[1], Liudmilla F. Kulikova[3], Valery A. Davydov[3], Viatcheslav N. Agafonov[4], Sergey I. Bozhevolnyi[1]*

[1]Centre of Nano Optics, University of Southern Denmark, DK-5230 Odense M, Denmark.

[2]Niels Bohr Institute, University of Copenhagen, 2100 Copenhagen, Denmark

[3]L.F. Vereshchagin Institute for High Pressure Physics, Russian Academy of Sciences, Troitsk, Moscow, 142190, Russia.

[4]GREMAN, CNRS, UMR 7347, INSA CVL, Université de Tours, 37200 Tours, France.

*Corresponding author. email: yinhui.kan@nbi.ku.dk; seib@mci.sdu.dk

**Abstract:**

Skyrmions, topologically stable field configurations, have recently emerged in classical optics as structured light for high-density data applications. Achieving controllable on-chip generation of single-photon skyrmions, while being highly desirable for quantum information technologies, remains challenging due to the nanoscale confinement of quantum emitters (QEs). Here we demonstrate a metasurface-integrated quantum emitter (metaQE) platform enabling room-temperature on-chip generation of single-photon skyrmions. Near-field coupling between QEs and propitiously designed surface arrays of meta-atoms mediates spin-orbit interaction, transforming nanoscale-localized dipole emission into free-propagating topologically structured photonic modes. By exploiting this approach for structuring quantum emission from different color centers in nanodiamonds, we realize diverse skyrmionic states, including high-order anti-skyrmions and skyrmionium, and thereby demonstrate its universality across QEs. Our work establishes a unified framework for on-chip structured quantum light sources, offering versatile control of high-dimensional topological states, such as skyrmions, and advancing scalable quantum photonic technologies.

**Main text:**

Skyrmions, first introduced in the 1960s as topological solitons in nonlinear field theory[1,2], have since been predicted and observed across diverse physical systems, including magnetic materials[3,4], condensates systems[5], and liquid crystals[6]. By virtue of their swirling textures and quantized topological charges, skyrmions enable robust information encoding in data storage, transfer, and processing[7–10]. In recent years, skyrmions have been extended into the optical domain, ranging from localized electromagnetic fields[11–13] to propagating vector beams with spatially varying polarization patterns[14–17]. The free-space skyrmionic beams have been demonstrated in tightly focused beam[18], optical fibers[19], and toroidal electromagnetic pulses[20], exhibiting unique features such as space-time nonseparability and topological resilience in disordered media or free-space propagation[21–23]. While classical optical skyrmions have been intensively studied, extending such topological configurations into the realm of quantum optics remains at an early stage.

Realizing skyrmionic textures at the single-photon level could unlock previously inaccessible degrees of freedom associated with spatially structured polarization, thereby enriching the landscape of high-dimensional quantum information processing and enhancing robustness against decoherence and environmental noise[24–26]. However, current implementations typically rely on complex optical setups, or require strict operational conditions, which hinder scalability and tunability[20,27–29]. These challenges make the on-chip generation of topologically structured quantum states particularly difficult, especially for nanoscale solid-state quantum emitters (QEs)[30]. The primary obstacle is achieving efficient light-matter interactions within confined geometries while simultaneously generating controllable topological features at the subwavelength scale. Therefore, overcoming these limitations necessitates a scalable approach that integrates quantum light emission with topological field control in a compact, chip-based platform.

In this work, we demonstrate the chip-scale generation of single-photon skyrmions using a metasurface-integrated quantum emitter (metaQE) platform. In our approach, light-matter interactions between the QE and the metasurface are mediated by nonradiative, QE-excited surface plasmon polaritons (SPPs), enabling a flexible control of spin-orbital coupling of single-photon emission. By engineering arrays of subwavelength meta-atoms, we realize different skyrmionic photon emissions from colour centres in nanodiamonds, thus establishing a versatile on-chip platform for the generation and exploration of topologically structured quantum light.

**Results**

**MetaQE for spin-orbit coupled quantum emission**

At-source manipulation of QE emission is inherently challenging due to QE nanoscale dimensions, thus leaving very limited space for shaping light-matter interactions[31]. To overcome this limitation, the developed metasurface-based approach enables nonradiative QE excitation of in-plane spreading surface electromagnetic waves, i.e., QE-excited SPPs[32]. A QE is located atop the 20 nm silica ($SiO_2$) and 150 nm silver (Ag) films and surrounded by an ultrathin, carefully designed and fabricated dielectric metasurface. Upon excitation, the QE generates an electromagnetic field primarily polarized normal to the surface, that couples efficiently to nonradiative SPPs propagating radially along the dielectric-metal interface[33]. Thus, the QE-excited SPPs serve as a mediator between the QE and the metasurface. With properly designed subwavelength meta-atoms of metasurface, the QE-excited SPPs are outcoupled to free-space radiation with tailored phase and polarization, thereby generating the desired skyrmionic photon emission (Fig. 1a).

The single-photon skyrmion is characterized by the spatially varying polarization states, where polarization is associated with a spin angular momentum (SAM) of photons and the spatial distribution originates from photons carrying an orbital angular momentum (OAM)[34]. The designed meta-QE generates a pair of locally entangled spin-orbit vectorial states corresponding to the poles of a higher-order Poincaré sphere (HOPS), which combines both SAM and OAM of the emitted single photons[25,26]. Consequently, the superposition of the two orthogonal states with distinct OAM produces the skyrmionic topology (Fig. 1b). To tailor the SAM, we employ the L-shaped meta-atoms oriented at $\pm 45°$ with respect to the SPP propagation direction (coloured by red and blue). The structural anisotropy enables selective outcoupling of the QE-excited SPPs into free-space photon emission with right (RCP) and left (LCP) circular polarizations (Fig. 1c). The OAM is characterized by the vortex topological charge $\ell$ that is related to the $2\pi\ell$ phase increment around the vortex core. To accurately manipulate the phase wavefront, the meta-atoms are arranged in various trajectories $C_{R(L)}$ (e.g. concentric or spiral) for accommodating the spatially varying phase of scattering fields, thereby generating photon emissions with Laguerre-Gaussian (LG) modes, $|LG_p^\ell\rangle$. The topological charge $\ell$ is determined by the number of spiral arms $m$, following the spin-orbit relation $\ell_{R(L)} = m \pm 1$[35,36]. Meanwhile, the radial index $p$ is dictated by the number of trajectories $C_{R(L)n}$ ($n = 1,2,3...$), representing how many annular rings occupied alternatively with the RCP(LCP)-type meta-atoms are used. Increasing $n$ introduces additional radial nodes in the field amplitude, corresponding to higher-order LG modes. In this way, both the azimuthal and radial mode orders of the emitted photons are controlled by the metasurface configuration, forming different LG modes with orthogonal polarizations. Consequently, the metaQE converts the localized QE emission into a free-space single-photon emission with a spatially varying polarization distribution.

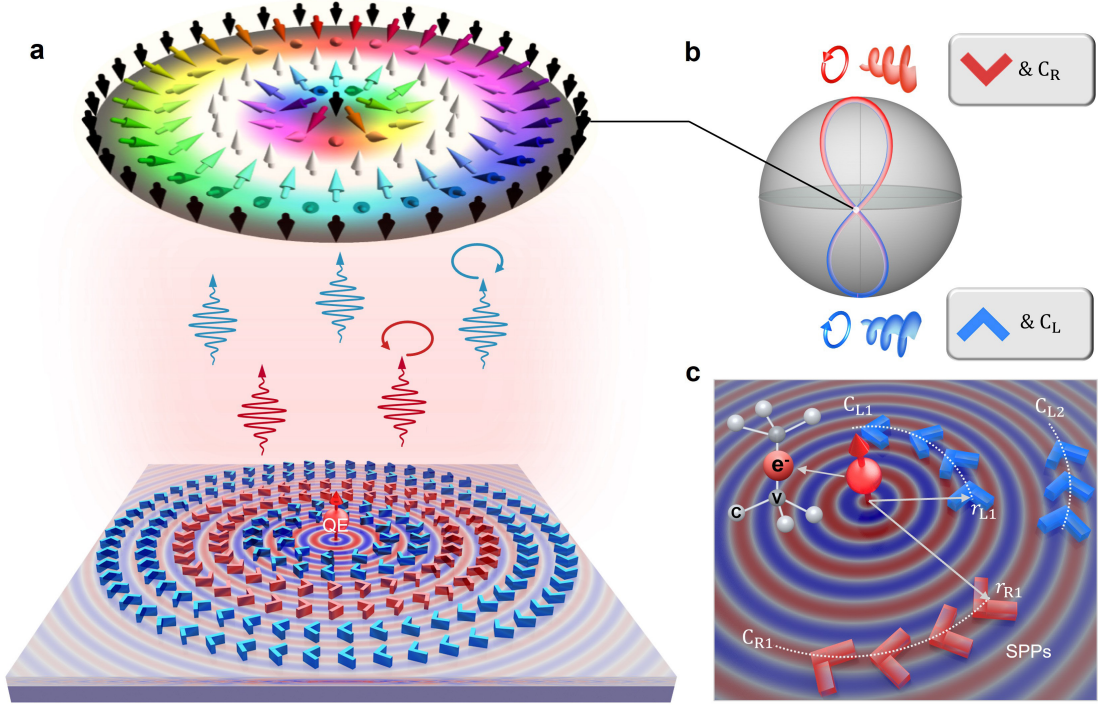

**Fig. 1 The schematic design of single-photon skyrmion emission from metaQE. a**, Illustration of metaQE emitting single photons with skyrmionium textures. Limited spatial dimensions of a nanoscale QE are extended by QE-excited SPPs, which are outcoupled by on-chip metasurface to free-space radiation with skyrmionic topology. The skyrmion texture is holistically visualized by the arrows showing Stokes vectors. **b**, With a strategically designed metasurface, the photon emission features orthogonal spatial and polarization modes, forming a pair of spin-orbit vectorial states represented on a HOPS. The two-mode superposition results in the skyrmionic photon emission. **c**, Design principle illustrating engineered the meta-atom chirality resulting in SAM, and azimuthal arrangement $C_{R(L)n}$ responsible for spatial mode and OAM. The QE features a radiative electrical dipole that is predominantly oriented normal to the surface, thereby efficiently exciting radially diverging, in-plane propagating SPPs. The red and blue meta-atoms with opposite chiralities outcouple the SPPs into RCP and LCP emission, respectively. $r_{R(L)}$ denotes the initial radii of the corresponding trajectories.

**Experimental realization of anti-type skyrmion**

Figure 2a shows metaQE design to construct skyrmionic photon emission with orthogonal polarizations and distinct LG modes. The two types of meta-atoms are arranged along Archimedean spiral trajectories: $r_{R(L)}(\varphi) = r_{R(L)1} + m\lambda_{SPP}\frac{\varphi}{2\pi}$, where $\varphi$ is the azimuthal angle of each meta-atom, $m$ is the number of spiral arms (here $m = 1$), $r_{R1}$ and $r_{L1}$ define the initial radii of the respective trajectories, and $\lambda_{SPP}$ is the propagation wavelength of SPPs (estimated by the filling ratio of metasurface and the QE emission wavelength[32]). Hence, combining the two metasurface can simultaneously modulate QE emission to produce $LG_0^0(\ell_R = 0, p = 0)$ mode with RCP and $LG_0^{-2}(\ell_L = -2, p = 0)$ mode with LCP, given by[29]:

$$|\Psi(r)\rangle = LG_0^0(r)|R\rangle + \exp(i\theta)LG_0^2(r)|L\rangle \qquad (1)$$

in which $\theta$ denotes the phase difference between the two modes, $|R\rangle$ and $|L\rangle$ represent RCP and LCP states. The spatial and azimuthal profile of OAM modes (inset of Fig. 2a) reveal that RCP component features a Gaussian beam with $\ell_R = 0$, whereas LCP possesses azimuthal phase of $\ell_L = -2$, forming a doughnut-shape intensity distribution. In experiment, the metasurface (made of HSQ, hydrogen silsesquioxane) was precisely integrated with a nanodiamond (ND) containing nitrogen-vacancy (NV) centres. The fabrication was facilitated by high-precision alignment markers and electron-beam lithography (EBL) techniques (see Supplementary Note S2 for details). The scanning electron microscopy (SEM) image of the device (Fig. 2b) shows that the meta-atoms have a width of 100 nm, a length of 400 nm, and are arranged with a fixed period of 580 nm ($\lambda_{SPP}$) along the radial direction. The false colour rendering highlights the opposite chirality of the meta-atoms, which form spirals with winding numbers of 2 and 7, respectively. The outer region is set with more winding numbers to compensate the attenuation of SPPs during their radial propagation. The structure ultimately forms nonseparable spin-orbital states at the pole of HOPS, thereby forming a synthesized optical skyrmion situated at the HOPS equator (Fig. 2c). The skyrmion topology is revealed by the polarization states characterized by the Stokes parameters $(S_0, S_1, S_2, S_3)$, in which the hue colour visualizes the azimuth of transverse $(S_1, S_2)$ component and lightness (from black to white) indicates the circular polarization component $S_3$ variation (from RCP to LCP). The topological property is characterized by the skyrmion number defined as:

$$N_{sk} = \frac{1}{4\pi}\iint \boldsymbol{S} \cdot \left(\frac{\partial \boldsymbol{S}}{\partial x} \times \frac{\partial \boldsymbol{S}}{\partial y}\right) dxdy \qquad (2)$$

with the Stokes vector $\boldsymbol{S} = (S_1, S_2, S_3)/S_0$, a nontrivial number $N_{sk}$ indicates the number of times the polarization vectors cover the full Poincare sphere. The obtained polarization field indicates the higher order anti-type skyrmion distributions with skyrmion number of -1.86, in close agreement with the ideal value $\Delta \ell = \ell_L - \ell_R = -2$. Moreover, another prominent feature of this design is that the skyrmion distribution can be independently controlled. The intensity and phase determine the latitude and longitude of HOPS, respectively. By varying the initial radii between the two spirals ($\Delta r = r_1 - r_2$), the relative phase $\theta$ between the two modes can be continuously tuned from 0 to $2\pi$ along the equator of HOPS, following $\theta = 2\pi\Delta r/\lambda_{SPP}$. This relative phase directly maps onto the initial phase $\gamma$ of the azimuthal vector distribution, enabling full coverage of all possible orientations of the helical field. The three representative cases shown in Fig. 2c illustrate this behaviour: choosing $\Delta r$ of $0, \frac{\pi}{4}\lambda_{SPP}$, and $\frac{\pi}{2}\lambda_{SPP}$, produces relative phase shifts of $0, \frac{\pi}{2}, \pi$, which results in rotation of the anti-skyrmion distribution of $\gamma = \pi$, $\frac{\pi}{2}$ and 0. For skyrmionic configurations with positive $N_{sk}$ (e.g. $\Delta \ell = 1$), tuning $\Delta r$ also allows a full transition between Néel-type ($\gamma = 0, \pi$) and Bloch-type ($\gamma = \pm\pi/2$) skyrmions, thereby providing complete control over the skyrmion helicity (Supplementary Note S1).

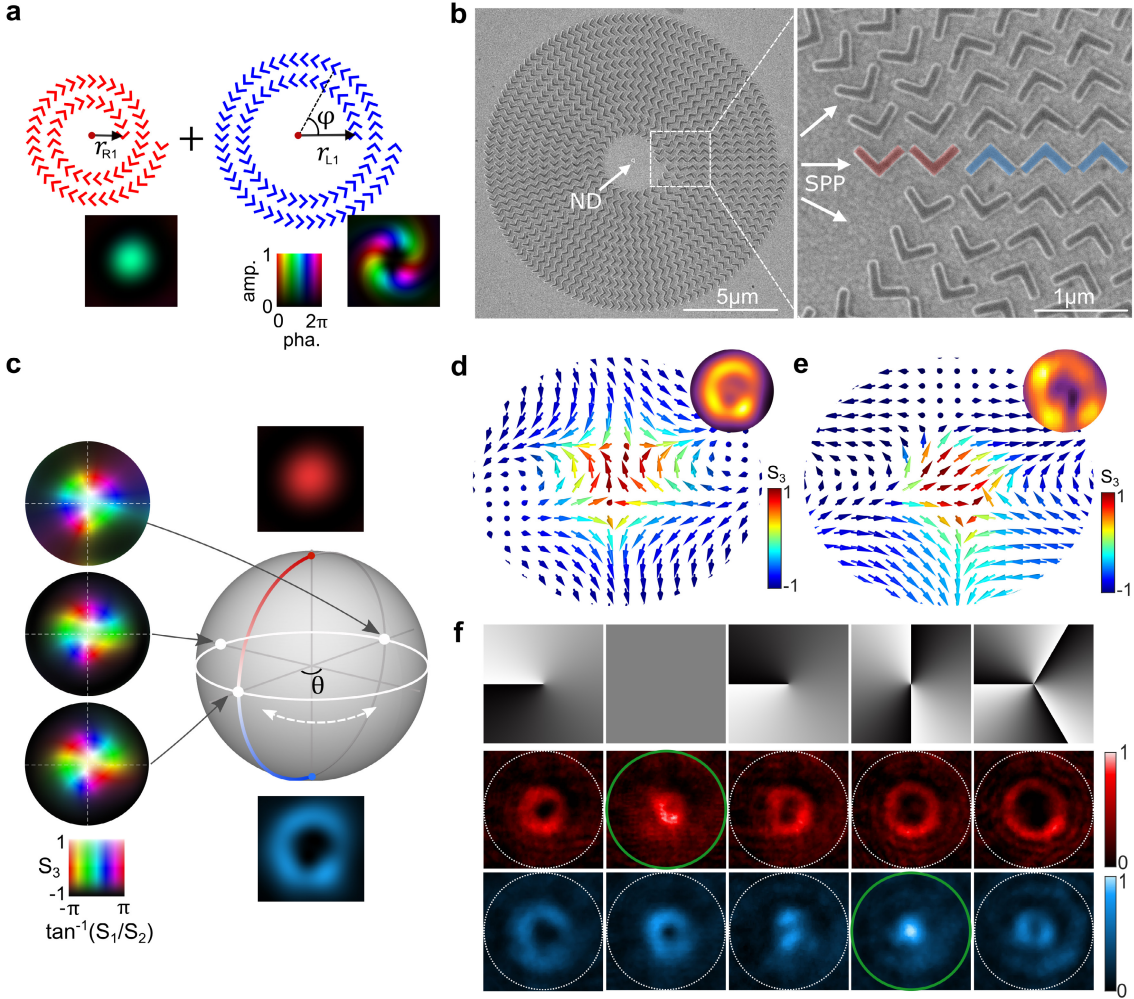

**Fig. 2 Anti-type skyrmion photon emission from NV centres in nanodiamond. a**, Metasurface design for funnelling QE emission to two eigenstates of LG mode. Meta-atoms of opposite chirality follow spiral trajectories with different initial radii, thereby generating RCP emission with $\ell_R = 0$ and LCP emission with $\ell_L = -2$. Inset: the far-field spatial phase distribution, the hue represents the phase, and the lightness denotes the intensity. **b**, SEM image of NV centres integrated with metasurface, where the magnified SEM image alongside highlights the varying chirality of meta-atom for controlling SAM. **c**, Evolution of skyrmionic emission with different helicities by tuning the initial radius. The poles of HOPS correspond to two eigenstates of LG mode, the angular coordinate $\theta$ denotes the relative phase between the modes controlled by the start radius. **d**, Simulated and **e**, Measured skyrmion map from the emitted photons, in which the vector arrows denote the spin azimuthal of Stokes vectors and the colour indicates circular polarized components $S_3$. **f**, Verification of the topological charge of RCP and LCP component of the photon emission. Polarization-resolved far-field emission patterns projected to SLM holograms with $\ell_{SLM} = -1, 0, 1, 2, 3$.

To efficiently excite the vertical dipole transition of the NV centres, we employed a 532 nm continuous-wave radially polarized laser, providing a dominant *z*-oriented electric-field component (experimental setup in Supplementary Note S3). The polarization-resolved far-field emission was then analysed by measuring the Stokes parameters using a quarter-wave plate and a linear polarizer (Supplementary Note S4). The skyrmion topology is illustrated by vectorial arrows, where the directions are derived from the Stokes parameters via $\arctan(S_1/S_2)$, and the colour varies with $S_3$. Both simulated (Fig. 2d) and measured (Fig. 2e) polarization states show a progressive change of the vector orientation from the 'up' state at the centre to the 'down' state at the edge, covering all intermediate polarization states from RCP to LCP. The corresponding spatial profile of the far-field emission, shown on the side, exhibits a divergence angle of 9°. Such a topology reveals the intrinsic intertwining between the spatial pattern and polarization, which is preserved even after the emitted photons pass through multiple optical components, including dichroic mirrors, reflection mirrors, bandpass filters, and imaging lenses. This preservation of the anti-skyrmion distribution, despite multiple reflections and optical transformations, demonstrates the topological protection of the emitted photons and their resilience to optical perturbations. To further analyse the distinct OAM modes, the RCP and LCP photon emissions were projected onto a spatial light modulator (SLM) encoded with LG phase holograms ($\ell_{\text{SLM}} = -1, 0, 1, 2, 3$). When the OAM of the emitted photons is opposite to that of the reference LG mode on the SLM, the diffracted beam appears as a Gaussian spot (Fig. 2f). The bright spots are observed at $\ell_{\text{SLM}} = 0$ for the RCP component and at $\ell_{\text{SLM}} = 2$ for the LCP component, verifying the topological charge of $\ell_R = 0$ and $\ell_L = -2$ of the photon emission.

**Engineering skyrmionium topology**

In general, the skyrmion topology collapses when the skyrmion number approaches zero, indicating a deformation or annihilation of the spin texture[17]. The concept of a skyrmionium refers to a bound pair of a skyrmion and an anti-skyrmion with opposite skyrmion numbers. Although its total skyrmion number is zero, the structure retains a nontrivial internal texture that enables stable, localized motion and suppresses the skyrmion Hall effect[37]. Despite its conceptual importance, the realization of such nontrivial topology in quantum light remains largely unexplored and has not yet been experimentally realized. Inspired by magnetic skyrmionium structures[38], we aim to engineer the photon emission of QEs to exhibit analogous topological textures in their polarization states, thereby generating propagating single photons carrying a skyrmionium texture. Achieving this requires a tailored coupling between SAM and OAM, which can be constructed by the superposition of certain eigenstate of $\text{LG}_0^1$ with no radial mode ($\ell_R = 1, p = 0$) and $\text{LG}_1^{-1}$ with radial mode ($\ell_L = -1, p = 1$). The different radial modes avoid overlapping spatial domains for RCP and LCP fields.

For coupling with the local QE, we consider the meta-atoms with opposite chirality arranged along concentric paths ($m = 0$) with different initial radii $r_{R1} = 2\lambda_{SPP}$, $r_{L1} = 4\lambda_{SPP}$, and $r_{L2} = 6.75\lambda_{SPP}$, as shown in Fig. 3a. The configuration generates three doughnut-shaped emission beams with distinct divergence angles, corresponding to controlled coupling of the QE-excited SPPs into multiple OAM channels. The spatial phase profile shows a clear spatial polarization separation and sequential transition of the dominant OAM states, from $|\ell_R = 1\rangle$ to $|\ell_L = -1\rangle$, and back to $|\ell_R = 1\rangle$ across the beam cross section, satisfying the conditions required for a skyrmionium polarization texture. Fig. 3b shows the SEM image of the fabricated metasurface integrated with a nanodiamond containing NV centres. The false colour rendering highlights the rotation of meta-atoms that control the SAM of the emitted photons, and the concentric starting radii tune the accessible range of OAM. The measured far-field intensity of RCP and LCP components reveal three well-separated doughnut-shaped beams with distinct divergence angles, in excellent agreement with numerical simulations (Fig. 3c, e). The spatial evolution of the polarization ellipses further displays a continuous transformation from RCP to LCP and back to RCP as the radial position increases (Fig. 3d, f), confirming the nested skyrmion and anti-skyrmion configuration characteristic of a skyrmionium. The measured net skyrmion number of 0.07 closely matches the simulated value of 0.05, with the slight deviation from zero attributed to non-uniform emission intensity and measurement noise. These results demonstrate a neutral yet topologically structured polarization field with a skyrmionium topology. Moreover, as discussed in Fig. 2c, varying the initial radii of the meta-atoms simply shifts the relative phase and thus rotates the skyrmionium azimuthally (Fig. S7).

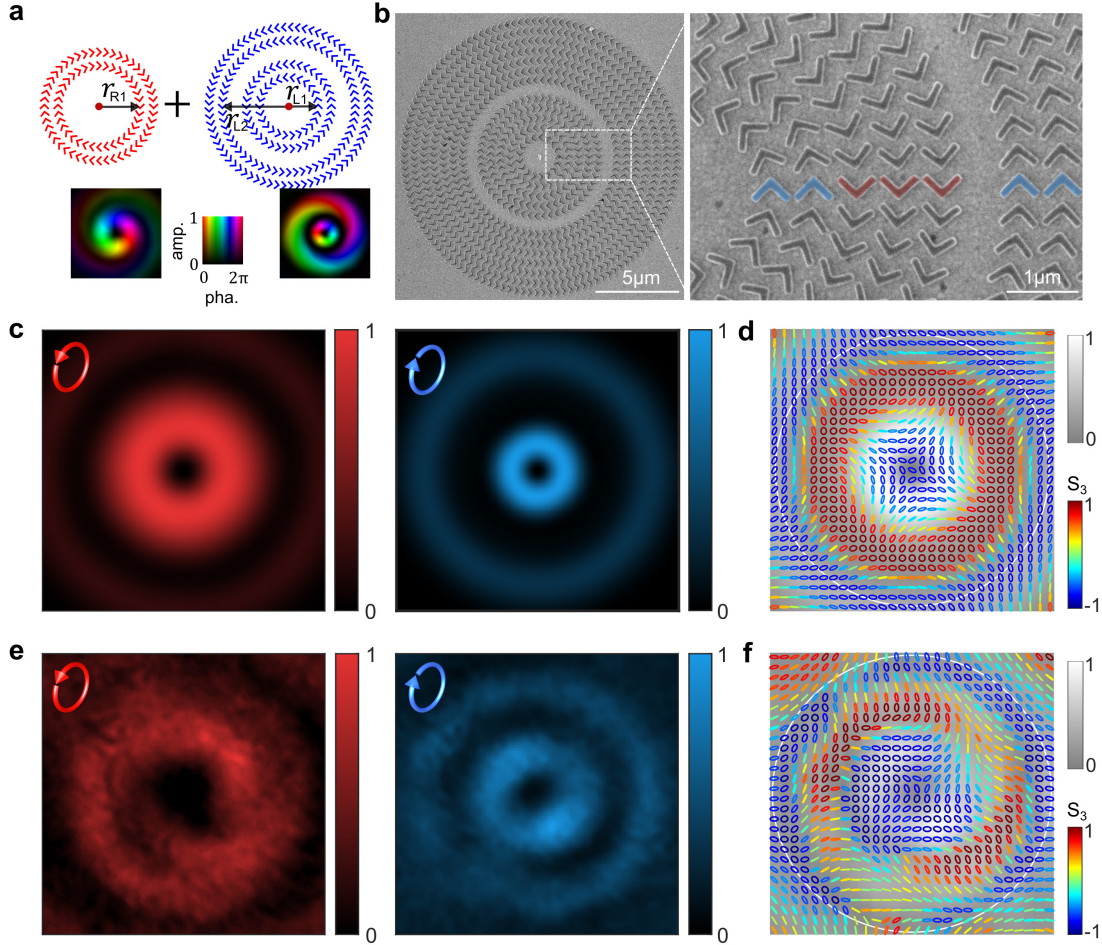

**Fig. 3 Realization of skyrmionium photon emission from NV centres. a,** Metasurface design for funneling QE emission to two eigenstates of $LG_0^1|R\rangle$ and $LG_1^{-1}|L\rangle$ mode. The meta-atoms are arranged along concentric paths with different starting radii, enabling spatial polarization separation and sequential transitions of OAM. **b,** SEM image of metasurface integrated with nanodiamonds hosting NV centres. **c,** Simulated and **e,** Measured far-field intensity profile of RCP and LCP component of the emitted photons. The displayed area corresponds to a NA of 0.2. **d,** Simulated and **f,** Measured polarization states superimposed on the emission intensity. The skyrmion regions are enclosed by white circles, and the ellipses denote the local polarization states, with colour representing the Stokes parameter $S_3$.

**Skyrmionium single-photon sources with GeV centres**

To demonstrate the generality of the approach, we applied the metasurface design to single germanium-vacancy (GeV) centres in nanodiamonds, which is particularly appealing for quantum technologies due to their high brightness, narrow linewidth, and exceptional spectral stability[39,40]. In the experiment, we preselect nanodiamonds hosting a single GeV centre with a dominant vertically oriented transition dipole[41]. The metasurface period was optimized to 490 nm ($\lambda_{SPP}$) to match the GeV-excited SPP mode. The photoluminescence spectrum of the device features a pronounced emission peak at 602 nm (Fig. 4a). The second-order correlation measurement using a Hanbury Brown-Twiss configuration yields $g^2(0) = 0.19$, confirming the single-photon emission at room temperature (Fig. 4b). The measured Stokes parameters show two successive SAM reversals, in excellent agreement with simulation (Fig. 4c, d). The measured polarization state further confirms the skyrmionium-type topology, exhibiting the characteristic pair of oppositely oriented spin-polarization lobes (Fig. 4e, f). The skyrmion number is 0.06 for the single-photon emission. The minor deviations and the slight rotation of the Stokes parameters arise mainly from asymmetries of the meta-atoms in fabrication, which lead to unequal RCP/LCP amplitudes and an intensity discontinuity at the lobe boundary (inset of Fig. 4f). Importantly, these results show that skyrmionium single-photon emission can be readily realized with different colour centres, and the underlying principle is fully compatible with other solid-state emitters, e.g. semiconductor quantum dots and defects in 2D materials. The metaQE design therefore provides a general and versatile route toward topologically structured single-photon sources.

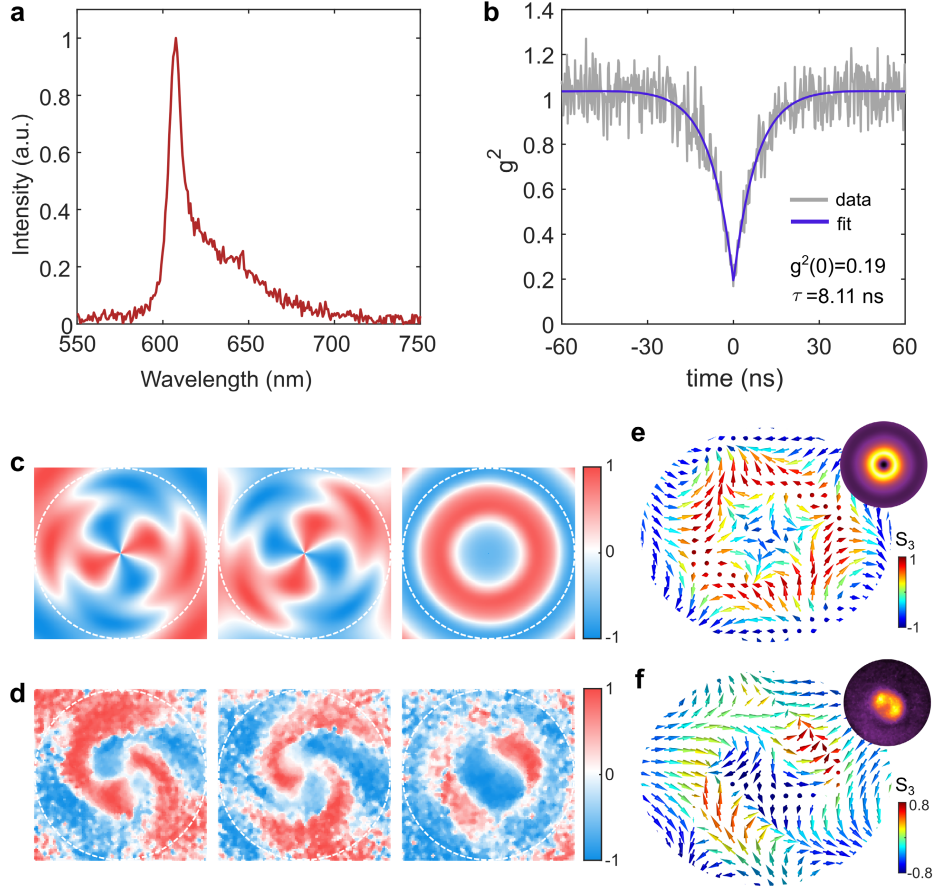

**Fig. 4 Single-photon skyrmionium emission from the GeV centre. a,** Photoluminescence spectrum from the metasruface integrated GeV centre. Under a continuous-wave excitation power of 100 μW, the detected photon rate reaches $7 \times 10^4$ counts/s. **b,** Second-order correlation $g^2(\tau)$ measured at room temperature. The lifetime is 8.11 ns. **c,** Simulated and **d,** Measured Stokes parameters of the single photon emission. **e,** Simulated and **f,** Measured polarized distribution of the emitted single photons, exhibiting a complete transition from RCP to LCP back to RCP. Inset: corresponding far-field intensity profiles.

## Conclusion

In summary, we have demonstrated a nanophotonic quantum light source that brings photonic topology into single-photon emission. By harnessing SPP-mediated light-matter interactions between individual QEs and engineered metasurfaces, we generate flying quantum skyrmions at room temperature, without requiring magnetic fields. The metaQE provides deterministic spin-orbit coupling at chip scale, facilitating the transfer of the dipole field into different structured topological modes. The resulting anti-skyrmion and skyrmionium photon emissions represent the first single-photon realizations of these polarization textures. The generality of the design, demonstrated across distinct colour centres, highlights the versatility and scalability of the metasurface approach. Looking ahead, combining our design with advances in the readout fidelity of colour centres could enable chip-scale quantum-enhanced sensing and biological imaging[42,43]. More broadly, our work establishes a pathway toward intrinsically robust quantum-information carriers and opens new opportunities for high-dimensional quantum state control.

## Methods

### Device fabrication

Our metaQE devices were fabricated following an established procedure (Fig. S4). A 150 nm Ag film was thermally evaporated onto a silicon substrate and coated with a 20 nm $SiO_2$ spacer by magnetron sputtering. Gold alignment markers were patterned by electron-beam lithography (EBL, JEOL-6490, 30 kV), metal deposition, and lift-off. Nanodiamonds containing colour centres (NV or GeV) were dispersed in deionized water at an optimized concentration and spincoated onto the prepared substrates to obtain spatially isolated NDs. Their positions were identified by dark-field microscopy referenced to the alignment markers. A 150 nm HSQ layer was then spin coated at 2200 rpm for 45 s and prebaked at 160 °C for 2 min. Subsequently, metasurfaces were written around selected NDs by EBL and developed in 25% TMAH for 4 min, followed by rinsing in deionized water.

### Optical characterization

Optical measurements were carried out using a home-built confocal micro-photoluminescence setup (Fig. S5). A radially polarized 532-nm laser beam was tightly focused onto the metaQE using a 100× objective (NA = 0.9, Olympus MPLFLN). The sample was scanned with a three-axis piezoelectric stage to identify the precise positions of QE. The emitted fluorescence was collected by the same objective, passed through a dichroic mirror (to filter the incident light), and detected by an avalanche photodiode (APD) for photon count mapping; Polarization-resolved emission was measured by inserting a quarter-wave plate and a linear polarizer before detection. Stokes parameters were obtained from sequential measurements with different analyser settings (Note S4); Second-order photon correlation measurements were performed using a Hanbury Brown-Twiss interferometer. The fluorescence was divided by a 50:50 non-polarizing beamsplitter and sent to two single-photon avalanche photodiodes. Photon arrival times were recorded using a time-correlated single-photon counting module (PicoHarp 300) operated in start-stop mode. The normalized coincidence histogram yielded the second-order correlation function $g^2(\tau)$, where $g^2(0) < 0.5$ confirmed single-photon emission. All measurements were carried out at room temperature.

## Numerical simulations

Three-dimensional finite-difference time-domain (FDTD) simulations were used to model the metaQE. A z-oriented electric dipole source was placed 30 nm above the $SiO_2$/Ag substrate and centered within the dielectric nanostructures (height 150 nm, refractive index 1.41). The dipole emission wavelength was set to 670 nm for NV centre and 602 nm for GeV centre to match their emission peaks at room temperature. Perfectly matched layer (PML) boundary conditions were applied in all directions. A nonuniform mesh was used with 10 nm resolution around the QE and the metal-dielectric interface. To extract the angular emission pattern, a two-dimensional field monitor was placed 30 nm above the metasurface to record the near-field distribution. The far-field radiation pattern was then obtained by applying a near to far field transformation, yielding the polarization-resolved far-field intensity distributions used for comparison with experimental measurements.

## Acknowledgements

The authors acknowledge the support from European Union's Horizon Europe research and innovation programme under the Marie Skłodowska-Curie Action (Grant agreement No. 101064471).

## Author contributions

Y.H.K. and S.I.B. conceived the idea. Y.H.K., X.J.L. and S.I.B. performed theoretical modelling. X.J.L. fabricated samples. X.J.L. and Y.H.K. performed experimental measurement. L.F.K., V.A.D., and V.N.A synthesized the GeV nanodiamonds. X.J.L., Y.H.K., S.K., and S.I.B. analysed the data. S.I.B. and Y.H.K. supervised the project. X.J.L. wrote the manuscript with contributions from all authors.

## Competing interests

The authors declare no competing interests.

## Reference

1. Skyrme, T. H. R. A non-linear feld theory. *Proc. R. Soc. A* **260**, 127–138 (1961).

2. Skyrme, T. H. R. A unified field theory of mesons and baryons. *Nucl. Phys.* **31**, 556–569 (1962).

3. Mühlbauer, S. *et al.* Skyrmion Lattice in a Chiral Magnet. *Science* **323**, 915–919 (2009).

4. Fert, A., Reyren, N. & Cros, V. Magnetic skyrmions: advances in physics and potential applications. *Nat. Rev. Mater.* **2**, 17031 (2017).

5. Al Khawaja, U. & Stoof, H. Skyrmions in a ferromagnetic Bose–Einstein condensate. *Nature* **411**, 918–920 (2001).

6. Foster, D. *et al.* Two-dimensional skyrmion bags in liquid crystals and ferromagnets. *Nat. Phys.* **15**, 655–659 (2019).

7. Yu, X. Z. *et al.* Real-space observation of a two-dimensional skyrmion crystal. *Nature* **465**, 901–904 (2010).

8. Nagaosa, N. & Tokura, Y. Topological properties and dynamics of magnetic skyrmions. *Nat. Nanotechnol.* **8**, 899–911 (2013).

9. Lima Fernandes, I., Blügel, S. & Lounis, S. Spin-orbit enabled all-electrical readout of chiral spin-textures. *Nat. Commun.* **13**, 1576 (2022).

10. Zheng, F. *et al.* Skyrmion–antiskyrmion pair creation and annihilation in a cubic chiral magnet. *Nat. Phys.* **18**, 863–868 (2022).

11. Du, L., Yang, A., Zayats, A. V. & Yuan, X. Deep-subwavelength features of photonic skyrmions in a confined electromagnetic field with orbital angular momentum. *Nat. Phys.* **15**, 650–654 (2019).

12. Dai, Y. *et al.* Plasmonic topological quasiparticle on the nanometre and femtosecond scales. *Nature* **588**, 616–619 (2020).

13. Tsesses, S. *et al.* Optical skyrmion lattice in evanescent electromagnetic fields. *Science* **361**, 993–996 (2018).

14. Gao, S. *et al.* Paraxial skyrmionic beams. *Phys. Rev. A* **102**, 053513 (2020).

15. Sugic, D. *et al.* Particle-like topologies in light. *Nat. Commun.* **12**, 6785 (2021).

16. Nape, I. *et al.* Revealing the invariance of vectorial structured light in complex media. *Nat. Photon.* **16**, 538–546 (2022).

17. Shen, Y. *et al.* Optical skyrmions and other topological quasiparticles of light. *Nat. Photon.* **18**, 15–25 (2024).

18. Ma, J., Xie, Z. & Yuan, X. Tailoring Arrays of Optical Stokes Skyrmions in Tightly Focused Beams. *Laser Photon. Rev.* **19**, 2401113 (2025).

19. He, T. *et al.* Optical skyrmions from metafibers with subwavelength features. *Nat. Commun.* **15**, (2024).

20. Shen, Y., Hou, Y., Papasimakis, N. & Zheludev, N. I. Supertoroidal light pulses as electromagnetic skyrmions propagating in free space. *Nat. Commun.* **12**, 5891 (2021).

21. Wang, R. *et al.* Observation of resilient propagation and free-space skyrmions in toroidal electromagnetic pulses. *Appl. Phys. Rev.* **11**, 031411 (2024).

22. Liu, C. Disorder-Induced Topological State Transition in the Optical Skyrmion Family. *Phys. Rev. Lett.* **129**, (2022).

23. Wang, A. A. *et al.* Perturbation-resilient integer arithmetic using optical skyrmions. *Nat. Photon.* (2025). doi:10.1038/s41566-025-01779-x.

24. Ornelas, P., Nape, I., de Mello Koch, R. & Forbes, A. Non-local skyrmions as topologically resilient quantum entangled states of light. *Nat. Photon.* **18**, 258–266 (2024).

25. He, C., Shen, Y. & Forbes, A. Towards higher-dimensional structured light. *Light Sci. Appl.* **11**, 205 (2022).

26. Kan, Y. *et al.* High-dimensional spin-orbital single-photon sources. *Sci. Adv.* **10**, eadq6298 (2024).

27. Zhang, Q. *et al.* Optical topological lattices of Bloch-type skyrmion and meron topologies. *Photon. Res.* **10**, 947 (2022).

28. Karnieli, A., Tsesses, S., Bartal, G. & Arie, A. Emulating spin transport with nonlinear optics, from high-order skyrmions to the topological Hall effect. *Nat. Commun.* **12**, 1092 (2021).

29. Shen, Y., Martı́nez, E. C. & Rosales-Guzmán, C. Generation of optical skyrmions with tunable topological textures. *ACS Photon.* **9**, 296–303 (2022).

30. Ma, J. *et al.* Nanophotonic quantum skyrmions enabled by semiconductor cavity quantum electrodynamics. *Nat. Phys.* **21**, 1462-1468 (2025).

31. Curto, A. G. *et al.* Unidirectional Emission of a Quantum Dot Coupled to a Nanoantenna. *Science* **329**, 930–933 (2010).

32. Kan, Y. *et al.* Metasurface-Enabled Generation of Circularly Polarized Single Photons. *Adv. Mater.* **32**, 1907832 (2020).

33. Liu, X. *et al.* Ultracompact Single-Photon Sources of Linearly Polarized Vortex Beams. *Adv. Mater.* **36**, 2304495 (2024).

34. Nape, I., Sephton, B., Ornelas, P., Moodley, C. & Forbes, A. Quantum structured light in high dimensions. *APL Photon.* **8**, 051101 (2023).

35. Wu, C. *et al.* Room-temperature on-chip orbital angular momentum single-photon sources. *Sci. Adv.* **8**, eabk3075 (2022).

36. Liu, X. *et al.* On-chip generation of single-photon circularly polarized single-mode vortex beams. *Sci. Adv.* **9**, eadh0725 (2023).

37. Göbel, B., Mertig, I. & Tretiakov, O. A. Beyond skyrmions: Review and perspectives of alternative magnetic quasiparticles. *Phys. Rep.* **895**, 1–28 (2021).

38. Zhang, X. *et al.* Control and manipulation of a magnetic skyrmionium in nanostructures. *Phys. Rev. B* **94**, 094420 (2016).

39. Senkalla, K., Genov, G., Metsch, M. H., Siyushev, P. & Jelezko, F. Germanium Vacancy in Diamond Quantum Memory Exceeding 20 ms. *Phys. Rev. Lett.* **132**, 026901 (2024).

40. Chen, D. *et al.* Quantum Interference of Resonance Fluorescence from Germanium-Vacancy Color Centers in Diamond. *Nano Lett.* **22**, 6306–6312 (2022).

41. Komisar, D. *et al.* Multiple channelling single-photon emission with scattering holography designed metasurfaces. *Nat. Commun.* **14**, 6253 (2023).

42. Wu, W. *et al.* Spin squeezing in an ensemble of nitrogen–vacancy centres in diamond. *Nature* **646**, 74–80 (2025).

43. Aslam, N. *et al.* Quantum sensors for biomedical applications. *Nat. Rev. Phys.* **5**, 157–169 (2023).

# Supplementary information

# Chip-integrated metasurface-enabled single-photon skyrmion sources

Xujing Liu[1], Yinhui Kan[1,2]*, Shailesh Kumar[1], Liudmilla F. Kulikova[3], Valery A. Davydov[3], Viatcheslav N. Agafonov[4], Sergey I. Bozhevolnyi[1]*

[1]Centre of Nano Optics, University of Southern Denmark, DK-5230 Odense M, Denmark.

[2]Niels Bohr Institute, University of Copenhagen, 2100 Copenhagen, Denmark

[3]L.F. Vereshchagin Institute for High Pressure Physics, Russian Academy of Sciences, Troitsk, Moscow, 142190, Russia.

[4]GREMAN, CNRS, UMR 7347, INSA CVL, Université de Tours, 37200 Tours, France.

*Corresponding author. email: yinhui.kan@nbi.ku.dk; seib@mci.sdu.dk

**Note S1. Generating skyrmionic beam**

The simplest form of a skyrmionic beam can be constructed by superposing two orthogonally polarized LG modes with no radial nodes ($p = 0$), the same beam width, a common focal point, whose OAM differ by one, as follows[1]:

$$|\Psi(r)\rangle = \text{LG}_0^0(r)|0\rangle + \exp(i\theta)\text{LG}_0^1(r)|1\rangle$$

Here, $|0\rangle$ and $|1\rangle$ represent any two orthogonal optical polarization (e.g. $|R\rangle$ and $|L\rangle$), while $\text{LG}_0^0(r)$ and $\text{LG}_0^1(r)$ correspond to two orthogonal spatial modes. The global phase difference between the two modes is denoted by $\theta$. In this case, the skyrmion number is $N_{sk} = \Delta\ell = \ell_L - \ell_R$, which is the difference of topological charge of the two modes.

The normalized fundamental LG beams ($p = 0$), whose total intensity integrates to unity, are expressed at the beam waists ($z - z_0 = 0$, near field) as[2]:

$$\text{LG}_0^\ell(r) = \|L_0^\ell(x) = 1\| = \sqrt{\frac{2}{\pi l!}} \cdot \frac{1}{w} \cdot \exp\left(-\frac{r^2}{w^2}\right) \cdot \left(\sqrt{2}\frac{r}{w}\right)^{|\ell|} \cdot \exp(i\ell\varphi)$$

where $w = w_0$ is the beam waist, $z$ denotes the propagation distance, $r$ is the radial distance from the beam centre, $\varphi$ is the azimuthal angle, and $\ell$ is the topological charge. The amplitude or intensity maximum occur at $r_{max} = w\sqrt{\frac{|\ell|}{2}}$. The LG modes used to construct skyrmionic beams are therefore:

$$\text{LG}_0^0 = \frac{\sqrt{2}}{w\sqrt{\pi}} \exp\left(-\frac{r^2}{w^2}\right)$$

$$\text{LG}_0^{\pm 1} = \frac{\sqrt{2}}{w\sqrt{\pi}} \exp\left(-\frac{r^2}{w^2}\right) \cdot \left(\sqrt{2}\frac{r}{w}\right) \cdot \exp(\pm i\varphi)$$

By superposing these modes with opposite circular polarizations, a skyrmionic optical field can be written in the near field as:

$$|\Psi(r, z = 0)\rangle = \frac{\sqrt{2}}{w\sqrt{\pi}} \exp\left(-\frac{r^2}{w^2}\right) \left\{\frac{1}{\sqrt{2}}\begin{pmatrix}1\\-i\end{pmatrix} + \exp(i\theta) \cdot \left(\sqrt{2}\frac{r}{w}\right) \cdot \exp(i\varphi) \cdot \frac{1}{\sqrt{2}}\begin{pmatrix}1\\i\end{pmatrix}\right\} =$$

$$= \frac{1}{w\sqrt{\pi}} \exp\left(-\frac{r^2}{w^2}\right) \left\{\begin{pmatrix}1\\-i\end{pmatrix} + \exp(i\theta) \cdot \left(\sqrt{2}\frac{r}{w}\right) \cdot \exp(i\varphi) \cdot \begin{pmatrix}1\\i\end{pmatrix}\right\}$$

In the far-field ($z - z_0 \to \infty$), the fundamental LG mode is expressed as:

$$LG_0^\ell(r, z \to \infty) = \sqrt{\frac{2}{\pi|\ell|!}} \cdot \frac{1}{\alpha_0 z} \cdot \exp\left(-\frac{\alpha^2}{\alpha_0^2}\right) \cdot \left(\sqrt{2}\frac{\alpha}{\alpha_0}\right)^{|\ell|} \cdot \exp(i\ell\varphi) \cdot \exp\left(-i\frac{\alpha^2 z}{\alpha_0 w_0}\right)$$

$$\cdot \exp\left[-i(|\ell| + 1)\frac{\pi}{2}\right]$$

where $\alpha = r/z$ is the diffraction angle and $\alpha_0 = \frac{\lambda}{\pi w_0}$ defines the beam divergence.

Fig. S1 demonstrates the calculated far-field intensity profiles of a skyrmionic optical field. The interference between the RCP and LCP components gives rise to a spatially structured polarization field with skyrmionic topology. By varying the relative phase difference $\theta$ between the two modes, the optical field evolves continuously between Néel- and Bloch-type skyrmions characterized by different helicity parameters $\gamma$. Furthermore, more versatile forms of optical skyrmions could be realized by combing LG modes with different topological charges. When the topological charge difference $\Delta\ell$ between the two modes is varied, distinct skyrmionic textures emerge. For example, $\Delta\ell = +1$ and $\Delta\ell = -1$ enables generation of Néel-type and anti-type skyrmions, respectively. $\Delta\ell = -2$ leads to the formation of a high-order anti-type skyrmion featuring a multi-lobed polarization structure and enhanced topological complexity. These results demonstrate that the skyrmion number and helicity of optical skyrmions can be flexibly tailored by the two LG modes.

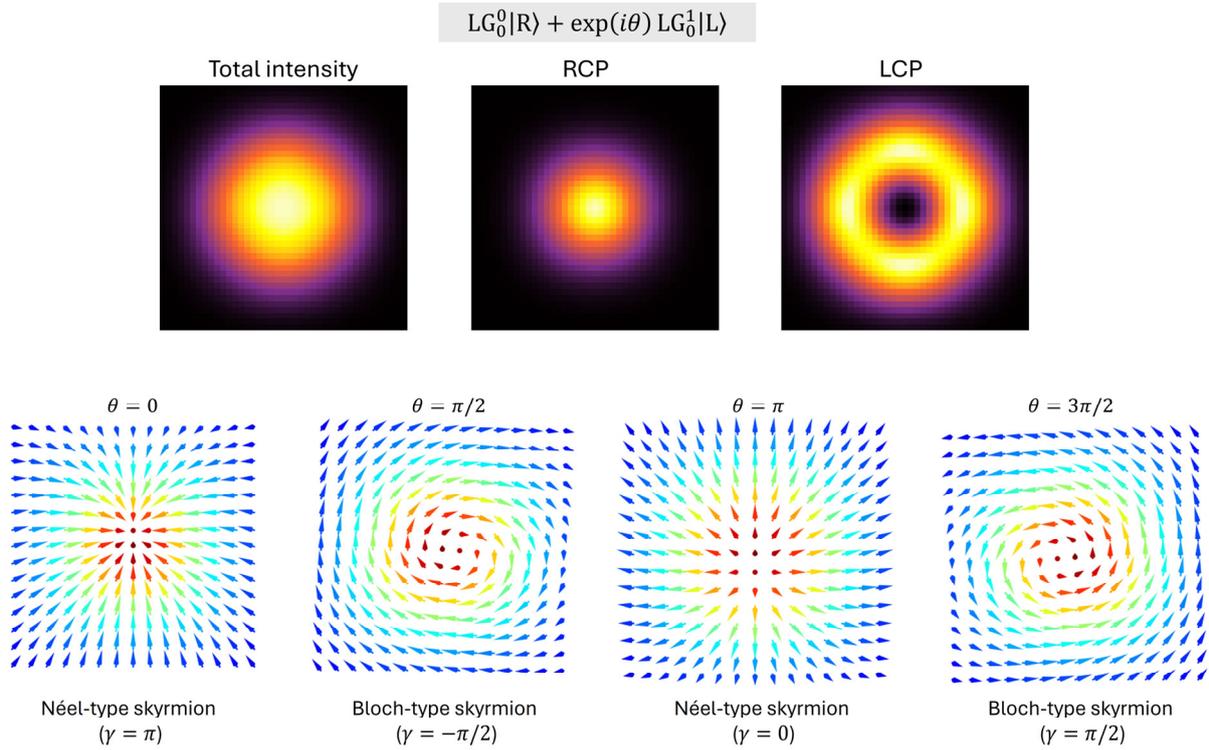

**Fig. S1** Far-field intensity profiles of a skyrmionic optical field by the superposition of two LG modes. Tuning the azimuthal phase difference $\theta$ between the modes leads to the continuous evolution between Néel- and Bloch-type skyrmions characterized by different helicity $\gamma$.

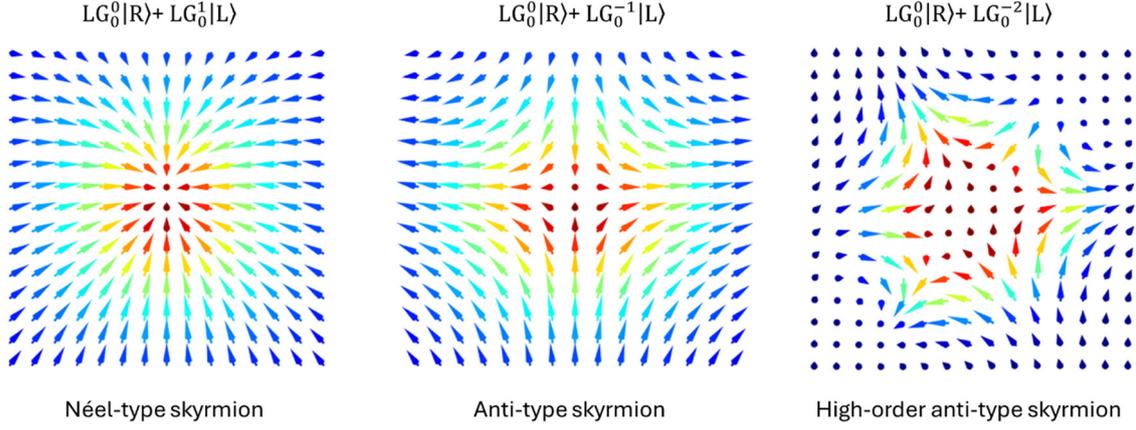

**Fig. S2** Polarization vector distributions of optical skyrmions formed by the interference between right- and left-circularly polarized LG modes with different topological charges.

For constructing skyrmionium beams, the fundamental LG beams with radial mode ($p = 1$) are used, which are described at their waists ($z - z_0 = 0$, near field) as follows:

$$\mathrm{LG}_1^\ell(r) = \left\|L_1^{|\ell|}(x) = 1\right\| = \sqrt{\frac{2}{\pi(1+|\ell|)!}} \cdot \frac{1}{w} \cdot \exp\left(-\frac{r^2}{w^2}\right) \cdot \left(\sqrt{2}\frac{r}{w}\right)^{|\ell|} \cdot \left(1 + |\ell| - \frac{2r^2}{w^2}\right) \exp(i\ell\varphi)$$

where $w = w_0$ is the beam waist, the field amplitude changes sign at the radius $r_{cross} = w\sqrt{\frac{1+|\ell|}{2}}$. Here, different radial modes for RCP and LCP fields are used to avoid spatial overlap between them. Thus, the skyrmionium field can be constructed by superposing $\mathrm{LG}_0^1$ and $\mathrm{LG}_1^{-1}$ modes with opposite circular polarizations. In the near field, this composite field generated by $\mathrm{LG}_0^1|R\rangle + \exp(i\theta)\,\mathrm{LG}_1^{-1}|L\rangle$ is written as:

$$|\Psi(r, z=0)\rangle = \frac{\sqrt{2}}{w\sqrt{\pi}} \exp\left(-\frac{r^2}{w^2}\right) \cdot \left(\sqrt{2}\frac{r}{w}\right) \cdot \exp(i\varphi) \cdot \frac{1}{\sqrt{2}}\binom{1}{-i} +$$

$$+ \exp(i\theta) \cdot \frac{1}{w\sqrt{\pi}} \exp\left(-\frac{r^2}{w^2}\right) \cdot \left(\sqrt{2}\frac{r}{w}\right) \cdot \left(2 - \frac{2r^2}{w^2}\right) \exp(-i\varphi) \cdot \frac{1}{\sqrt{2}}\binom{1}{i} =$$

$$= \frac{1}{w\sqrt{\pi}} \exp\left(-\frac{r^2}{w^2}\right) \cdot \left(\sqrt{2}\frac{r}{w}\right) \left\{\exp(i\varphi) \cdot \binom{1}{-i} + \exp(i\theta) \cdot \left(2 - \frac{2r^2}{w^2}\right) \cdot \exp(-i\varphi) \cdot \frac{1}{\sqrt{2}}\binom{1}{i}\right\}$$

Figure S3 shows the intensity and polarization distributions of optical skyrmionium fields. As $\theta$ varies from 0, $\pi/2$ to $\pi$, the skyrmionium field exhibit continuous rotation and internal twisting, corresponding to different helicity parameters ($\gamma = 0, \pi/2, \pi$).

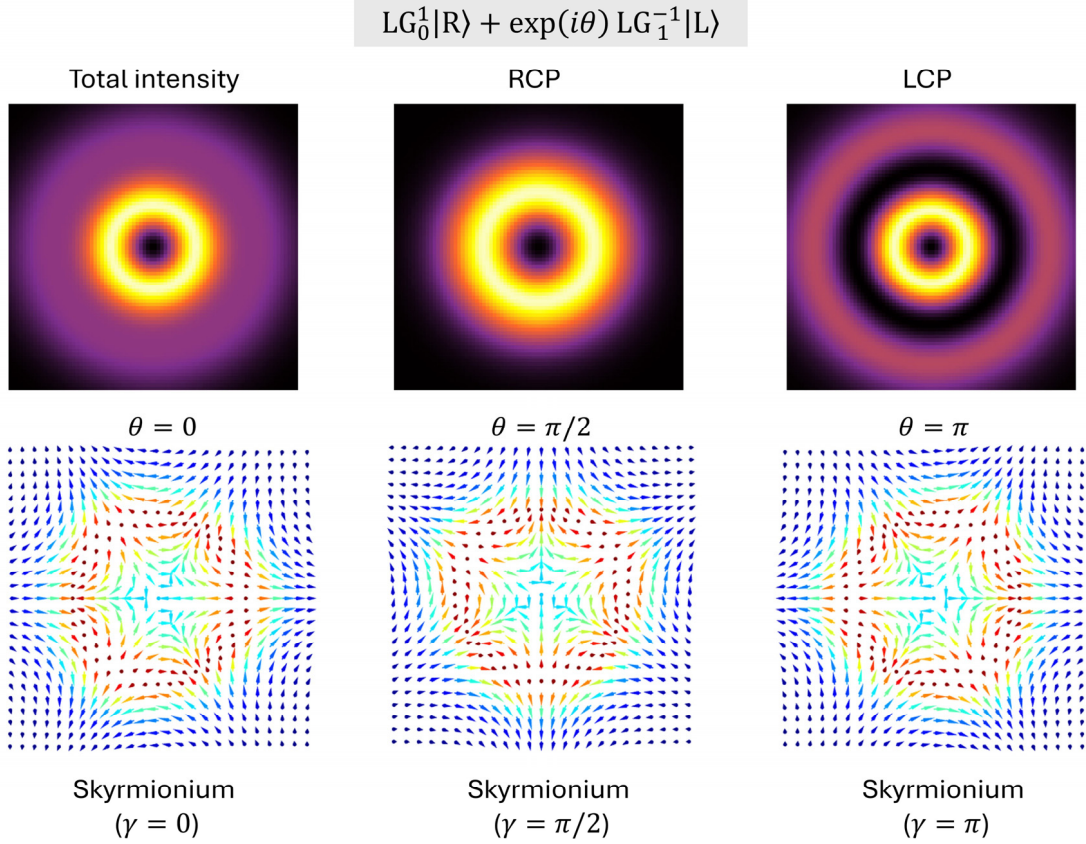

**Fig. S3** Intensity profiles of a skyrmionium optical field. Tuning the relative phase θ between the two modes makes a *skyrmionium* configuration helicity.

**Note S2 Device fabrication**

The fabrication workflow of the metasurface-integrated quantum emitters is summarized in Fig. S4. The fabrication process of metaQE is illustrated as follows. A 150 nm-thick Ag film was first thermally evaporated onto a silicon substrate, followed by deposition of a 20 nm $SiO_2$ layer using magnetron sputtering. Arrays of gold alignment markers were then defined on the substrate by electron-beam lithography (EBL, JEOL-6490, accelerating voltage 30 kV), gold deposition, and a standard lift-off process. A solution of nanodiamonds (NDs) containing colour centres was prepared at an optimized concentration and subsequently spin-coated onto the prepared substrates to obtain well-isolated NDs. The positions of the NDs were identified using dark-field microscopy image, with reference to the alignment markers. After determining the ND positions, a negative resist layer of hydrogen silsesquioxane (HSQ) was spin-coated at 4000 rpm for 45 s and prebaked at 160 °C for 2 min, forming a ~150 nm-thick layer verified by atomic force microscopy (NT-MDT NTEGRA). Metasurface structures was then carried out around the selected NV–NDs by EBL. The final metaQE devices were obtained after development in 25% tetramethylammonium hydroxide (TMAH) for 4 min, followed by rinsing in isopropanol for 60 s.

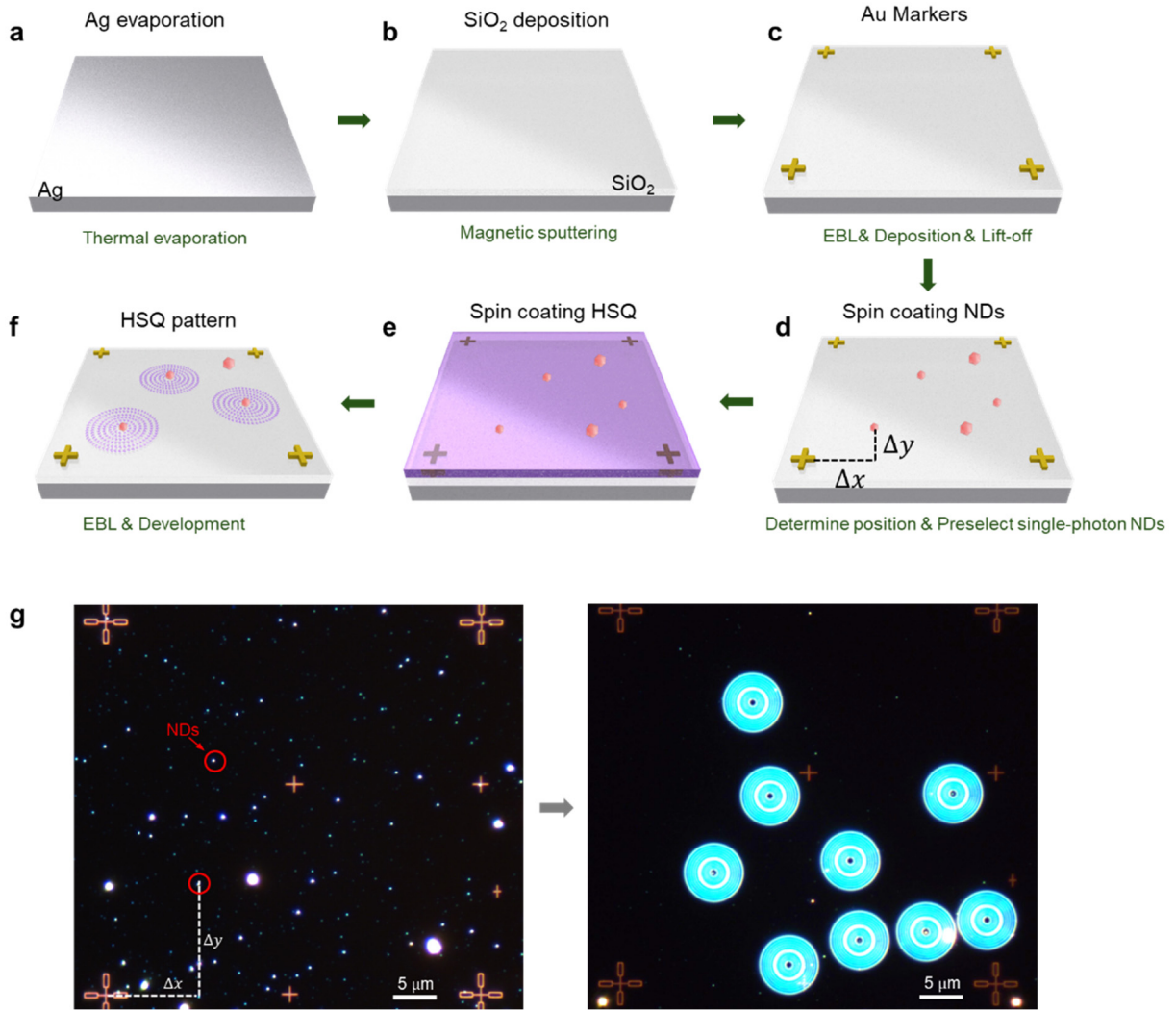

**Fig. S4** Schematic illustration of the fabrication process of meta-QE devices. (a) Thermal evaporation of a 150 nm Ag film on a Si substrate. (b) Deposition of a 20 nm SiO$_2$ layer by magnetic sputtering to form the metal–dielectric base. (c) Gold alignment markers EBL, gold deposition, and lift-off. (d) Spin-coating of NDs, the positions are determined with respect to the alignment markers. (e) Spin-coating of an HSQ resist layer for EBL patterning, followed by alignment of the metasurface pattern to the pre-selected NDs. (f) EBL writing and development. (g) Dark-field image of spin-coated NDs before and after coupled with metasurface.

**Note S3 Optical Characterization**

Optical measurements were performed using a confocal micro-photoluminescence setup, as shown in Fig. S5. The NDs were optically excited by a radially polarized laser beam focused onto the sample through a 100×, NA = 0.9 objective lens. 0 The fluorescence photon count rate was recorded using an avalanche photodiode (APD1) while scanning the sample with a piezoelectric stage, producing fluorescence maps for locating individual QEs. Polarization-resolved measurements were performed by inserting a broadband quarter-wave plate mounted on a motorized rotation stage and a linear polarizer into the detection path. A flip mirror was used to redirect the collected emission to a spatial light modulator (SLM) for characterization of the topological charge. The SLM was programmed with computer-generated phase holograms corresponding to different orbital angular momentum (OAM) modes. By projecting the emitted field onto these holograms and recording the first-order diffraction patterns on a CMOS2 camera, the topological charge of the structured emission was determined from the intensity distribution (Fig. 2e). Second-order correlation measurements were performed to confirm the single-photon emission. Second-order correlation measurements were carried out in a Hanbury Brown-Twiss (HBT) configuration to verify the quantum nature of the emitted photons (Fig. S6). The collected fluorescence was split by a 50:50 beam splitter into two detection channels connected to avalanche photodiodes (APD1 and APD2). Photon arrival times were recorded by a time-correlated single-photon counting (TCSPC) module (PicoHarp 300), operating in a start-stop mode. The coincidence histogram of detection events as a function of the delay time $\tau$ yielded the $g^2(\tau)$ function, which was normalized to unity at long delays. A dip below 0.5 at $\tau = 0$ confirmed the single-photon emission from individual colour centres.

**Note S4 Measurement of spatial polarization states and polarization ellipse**

The Stokes vector $\mathbf{S} = [S_0, S_1, S_2, S_3]$ is used to describe the degree of polarization for any state of light. $S_0$ represents the total intensity, $S_1$ and $S_2$ quantify the degree of linear polarization, and $S_3$ describes the degree of circular polarization. For a plane wave with orthogonal components $\mathbf{E_x} = A_x e^{i\varphi_x}$, $\mathbf{E_y} = A_y e^{i\varphi_y}$. The Stokes parameters are defined as:

$$S_0 = \mathbf{E_x E_x^*} + \mathbf{E_y E_y^*} = |A_x|^2 + |A_y|^2$$

$$S_1 = \mathbf{E_x E_x^*} - \mathbf{E_y E_y^*} = |A_x|^2 - |A_y|^2$$

$$S_2 = \mathbf{E_x E_y^*} + \mathbf{E_y E_x^*} = 2|A_x||A_y|\cos\delta$$

$$S_3 = i(\mathbf{E_x E_x^*} + \mathbf{E_y E_y^*}) = 2|A_x||A_y|\sin\delta$$

where $\mathbf{E_x^*}$ and $\mathbf{E_y^*}$ are the complex conjugates of $\mathbf{E_x}$ and $\mathbf{E_y}$, respectively, and where $\delta = \varphi_x - \varphi_y$ is the phase difference between the two field components.

To experimentally characterize the spatially varying polarization states of the emitted photons, we performed full Stokes polarimetry based on intensity quantities. The emitted photons pass through a quarter wave plate (retardation angle $\delta$). Then the beam passes a linear polarizer with its transmission axis aligned at an angle $\alpha$ to the $x$- axis. The far-field image was recorded for four combinations of $\delta$ and $\alpha$ to calculate the four Stokes parameters. The first three Stokes parameters are measured by rotating the linear polarizer to angle $\alpha = 0°, 90°, 45°,$ and $-45°$ respectively (without the quarter wave plate). The final parameter $S_3$ is measured using linear polarizer and the quarter wave retarder ($\alpha = \pm 45°, \delta = 90°$). The Stokes parameters are derived as:

$$S_0 = I(0°, 0°) + I(90°, 0°)$$
$$S_1 = I(0°, 0°) - I(90°, 0°)$$
$$S_2 = I(45°, 0°) - I(-45°, 0°)$$
$$S_3 = I(45°, 90°) - I(-45°, 90°)$$

The normalized Stokes parameters $\boldsymbol{s} = [S_0, S_1, S_2, S_3]/S_0$ were subsequently employed to calculate the local polarization ellipse. The shape and handedness of the ellipse are characterized by the ellipticity angle $\chi$ and the orientation angle $\psi$, which can be derived from the normalized Stokes parameters as

$$\sin 2\chi = S_3; \ \tan 2\psi = \frac{S_2}{S_1}$$

Here, $\psi$ describes the rotation of the polarization ellipse with respect to $x$-axis, and $\chi$ quantifies its ellipticity, where $\chi = 0°$ corresponds to linear polarization and $\chi = \pm 45°$ corresponds to right or lefthanded circular polarization. By mapping $\psi$ and $\chi$ across the beam profile, the local polarization ellipses can be reconstructed, providing an intuitive visualization of the spatially varying polarization states and their topological organization in skyrmionic and skyrmionium beams. The reconstructed polarization ellipses provide a direct visualization of the spatial polarization topology. To quantify the topological nature of the polarization field, we calculate the skyrmion number $N_{sk}$, which represents the number of times the normalized Stokes vector wraps around the Poincaré sphere. A nonzero $N_{sk}$ indicates a topologically nontrivial polarization configuration. For instance, a value close to $+1$ or $-1$ corresponds to a skyrmion or anti-skyrmion, while a near-zero value with an internal spin reversal indicates a skyrmionium structure.

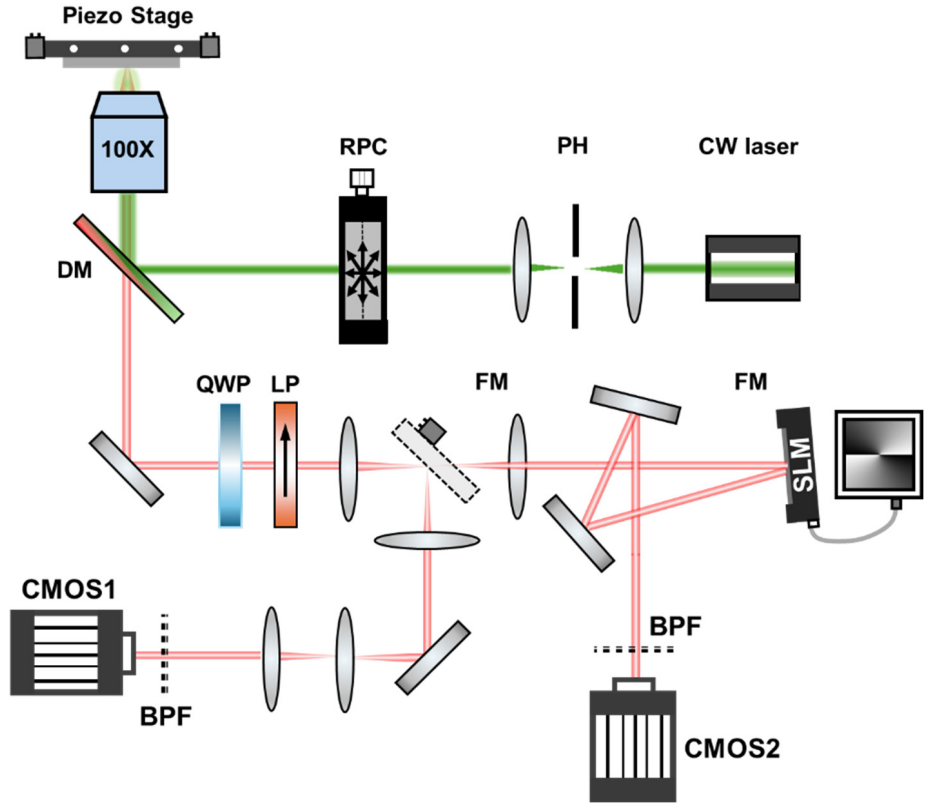

**Fig. S5** Schematic of the optical setup used for characterizing the meta-QE emission. The excitation beam is radially polarized and focused onto the sample through a 100×, NA = 0.9 objective. the spatially structured radiation pattern is captured at Fourier-plane images. CW: continuous wave, RP: Radial Polarization Converter, PBS: polarized beam splitter, PH: pinhole, DM: dichroic mirror, LP: linear polarization, QWP: quarter-wave plate, LPF: 550 nm long pass filter, FM: flip mirror.

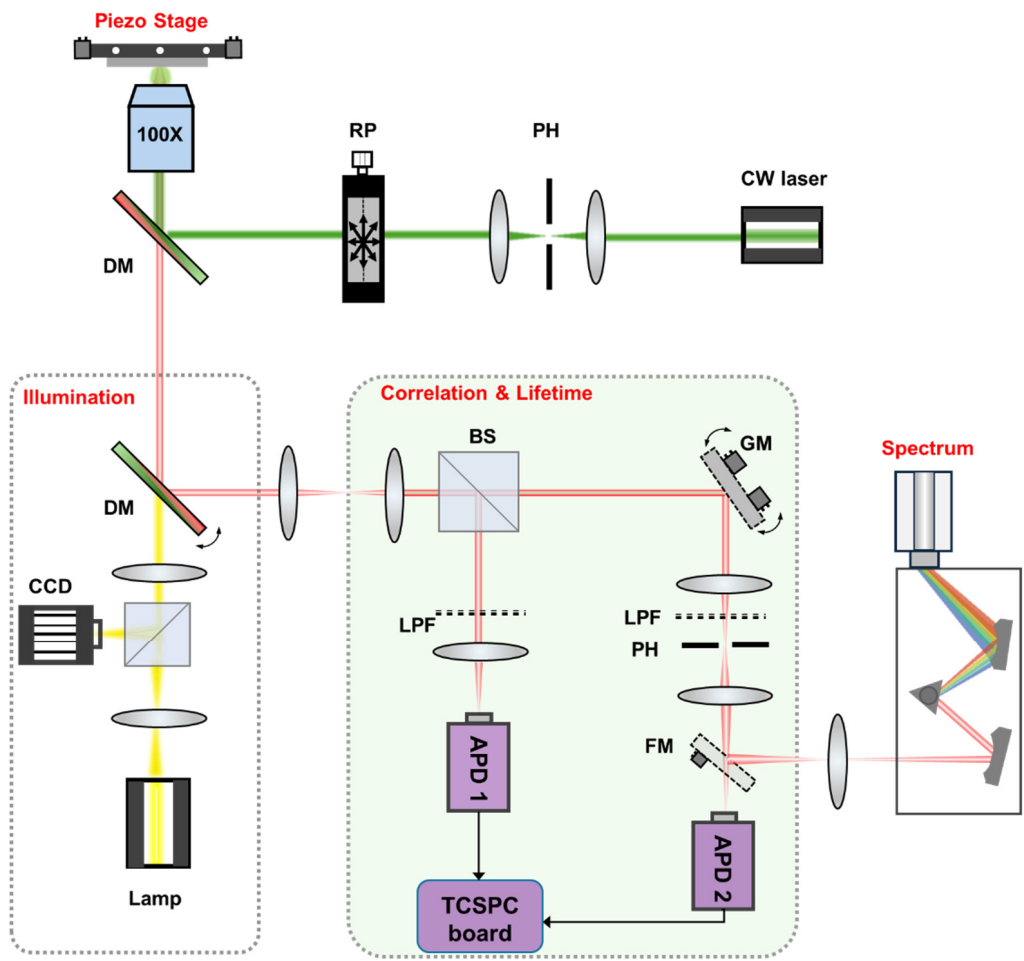

**Fig. S6** Experimental setup for optical characterization for correlation, lifetime, and spectral measurements. GM: galvanometric mirror. APD: Avalanche Photodiode.

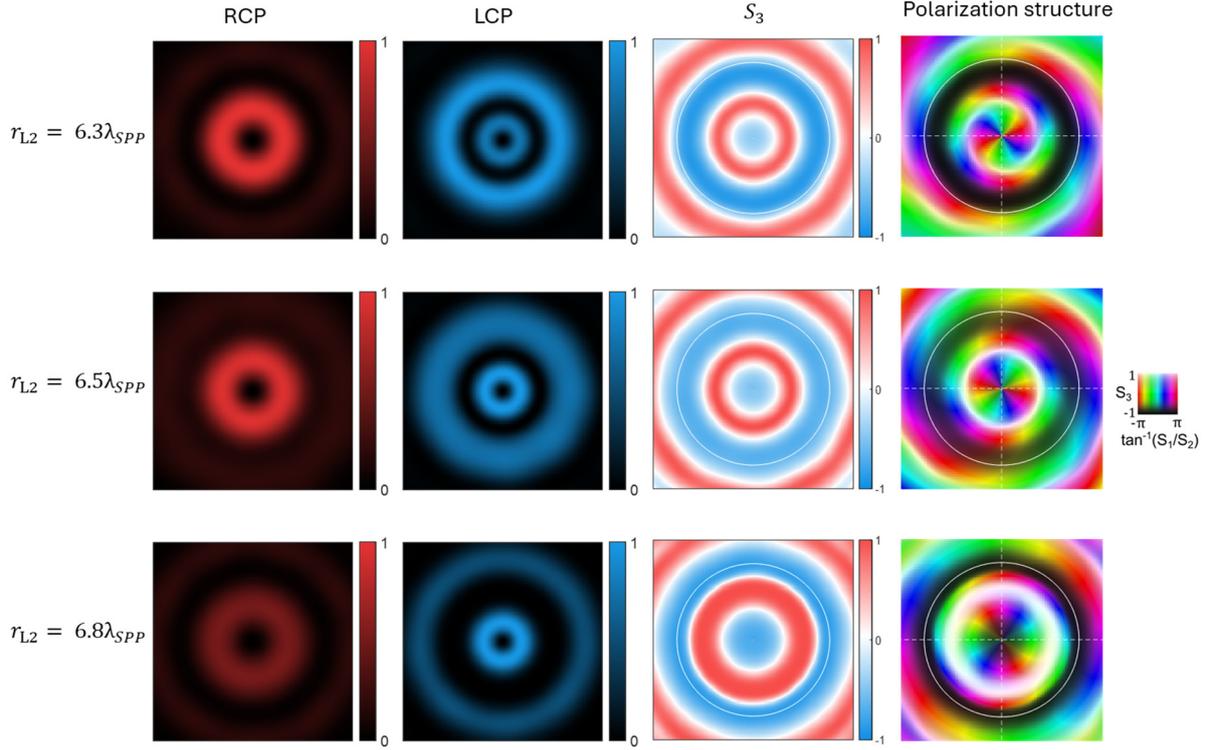

**Fig. S7** Simulated far-field distributions of photon emissions for three metasurface configurations with different initial radii $r_{L2}$. From left to right: intensity of the RCP and LCP components, the Stokes parameter $S_3$, and the spatial polarization state distribution. The sequential variation in the velocity of the polarization field verifies the tunable skyrmionium-like topology governed by the metasurface design, where increasing the initial radius of the meta-atoms systematically modifies the azimuthal phase accumulation.

**Reference:**

1. Gao, S. *et al.* Paraxial skyrmionic beams. *Phys. Rev. A* **102**, 053513 (2020).
2. Nape, I., Sephton, B., Ornelas, P., Moodley, C. & Forbes, A. Quantum structured light in high dimensions. *APL Photon.* **8**, 051101 (2023).